\documentclass{PoS}

\title{Continuum EoS for QCD with $N_f=2+1$ flavors}

\ShortTitle{Continuum EoS for QCD}

\author{
Szabolcs~Bors\'{a}nyi$^a$,
Zolt\'{a}n~Fodor$^{a,b,c}$,
Christian~Hoelbling$^{a}$, 
S\'{a}ndor~D.~Katz$^{c,d}$,
\speaker{Stefan Krieg$^{a,b}$},
K\'alm\'an~K.~Szab\'o$^{a,e}$\\
\ \\
$^a$Department of Physics, University of Wuppertal,\\ Gau\ss str. 20, D-42119, 
Germany\\
$^b$Forschungszentrum J\"ulich, J\"ulich, D-52425, Germany\\
$^c$Institute for Theoretical Physics, E\"otv\"os University,\\ P\'azm\'any
1, H-1117 Budapest, Hungary\\
$^d$MTA-ELTE Lend\"ulet Lattice Gauge Theory Research Group\\
$^e$Institute for Theoretical Physics, Universit\"at Regensburg\\
D-93040 Regensburg, Germany
}

\abstract{We report on a continuum extrapolated result~\cite{Borsanyi:2013bia} for the equation of state (EoS) of QCD with $N_f=2+1$ dynamical quark flavors. In this study, all systematics are controlled, quark masses are set to their physical values, and the continuum limit is taken using at least three lattice spacings corresponding to temporal extents up to $N_t=16$. A Symanzik improved gauge and stout-link improved staggered fermion action is used. Our results are available online~\cite{resultsonline}.}

\FullConference{31st International Symposium on Lattice Field Theory - LATTICE 2013\\
		July 29 - August 3, 2013\\
		Mainz, Germany}

\begin{document}

\section{Introduction}
The rapid transition from the quark-gluon-plasma 'phase'\footnote{Since this transition is a cross-over~\cite{Aoki:2006we}, this use of the term 'phase' is somewhat abusive, and indicates only the dominant degrees of freedom.} to the hadronic phase in the early universe and the QCD phase diagram are subjects of intense study in present heavy-ion experiments (LHC@CERN, RHIC@BNL, and the upcoming FAIR@GSI). This transition can be studied in a systematic way in Lattice QCD (for recent reviews see, e.g., \cite{Fodor:2009ax,Philipsen:2012nu,Petreczky:2013qj}). The associated (pseudo-)critical temperature scale $T_c$ is, given that the transition is a cross-over~\cite{Aoki:2006we}, not uniquely defined but depends on the observable considered. For any given observable, $T_c$ has, however, a well-defined value. 

The first full result\footnote{Here, we use the expression 'full' to indicate that a calculation used physical quark masses and included a controlled continuum extrapolation.} for $T_c$ from 2006~\cite{Aoki:2006br} was confirmed by later simulations including successively finer lattice spacings~\cite{Aoki:2009sc,Borsanyi:2010bp} and independent calculations using a different action~\cite{Bazavov:2011nk}. The now accepted value for the chiral transition is $T_c=150$~MeV (depending on the exact definition of the observables, for other observables see~\cite{Borsanyi:2010bp}). These results were extended to small baryonic chemical potentials ($\mu_B$)~\cite{Endrodi:2011gv,Borsanyi:2012cr} by means of the multiparameter-reweighting method of ref.~\cite{Fodor:2001au}. These (full) results provide the curvature of the phase diagram in the T-$\mu_B$ plane.

The equation of state (EoS) of QCD, (i.e, the pressure $p$, energy density $\epsilon$, trace anomaly $I=\epsilon-3p$, entropy $s=(\epsilon+p)/T$, and the speed of sound $c_s^2=dp/d\epsilon$ as functions of the temperature) has been determined by several groups, however, a full result was still lacking. Past calculations by the hotQCD Collaboration (p4, asqtad and hisq actions with $N_t$=6, 8, 10, and 12) resulted in a peak height of 5-8 for the peak of the trace anomaly ($I/T^4$, for a recent summary see ref.~\cite{Petreczky:2012fsa}), whereas our past results (Wuppertal-Budapest collaboration, 'WB', using the stout improved action), from 2005 onwards consistently showed a peak height of about 4~\cite{Borsanyi:2010cj}. The results of ref.~\cite{Borsanyi:2010cj} constitute a full result at three characteristic temperatures, which we now extended to the full temperature range~\cite{Borsanyi:2013bia}. This is the work discussed in these proceedings. Concerning the peak height, our calculations confirm our earlier findings, leaving a resolution of the discrepancy for, hopefully, Lattice 2014. For readers interested in using our results for the EoS, we made them available electronically~\cite{resultsonline}.

\section{Action and simulation setup}
Our calculation is based on a tree-level Symanzik improved gauge action with 2-step 
stout-link improved staggered fermions. The precise definition of the 
action can be found in ref.~\cite{Aoki:2005vt}, its advantageous scaling properties are discussed in ref.~\cite{Aoki:2009sc}. In particular, while it approaches the continuum value of the Stefan-Boltzmann limit in the infinite temperature limit $T\rightarrow\infty$ slower than actions with p4 or Naik terms (the latter is an additional fermionic term in the asqtad and hisq actions), it behaves monotonous and reaches the asymptotic $a^2$ behavior quite ``early''. Extrapolations from moderate temporal 
extents, e.g., using $N_t \ge 8$, allow for a smooth continuum extrapolation and provide an accuracy on the percent level, the typical accuracy one aims to reach. Additionally, applying simple tree-level improvement factors for the bulk thermodynamic observables brings the individual data points for the different $N_t$ very close to the continuum limit. Since a simulation with our action requires much less computational resources, we can have several lattice spacings, enabling us to take a controlled continuum extrapolation. Other improved actions, which are less local, such as p4 or the Naik-type asqtad/HISQ, can have non-monotonic behavior (consider, e.g., the $N_t$ dependence of the free energy density for the Naik term)~\cite{Peikert:1997vf}. In $T=0$ simulations, which enter the $T>0$ data through renormalization and scale setting, these actions still have $O(a^2)$ cutoff effects. Therefore, their improved scaling at $T\rightarrow\infty$ cannot remove all $O(a^2)$ lattice artifacts.

Taste-breaking artifacts (for an extended discussion see~\cite{Borsanyi:2013bia}), which turned out to be more important than an improved $T\rightarrow\infty$ limit, are effectively reduced by gauge smearing, which motivated our selection of the 2-stout action (see Figure~1 of ref.~\cite{Aoki:2005vt} or Figure~2 of ref.~\cite{Borsanyi:2010bp}). As of today, the new HISQ action possesses an even smaller taste violation (see, e.g., Figure 4 of ref.~\cite{Bazavov:2011nk}), though at higher computational costs. Small taste breaking artifacts improve the precision of the continuum limit particularly at low temperatures, where lattice spacings are coarse. As long as the lattice spacings used are within in the scaling window, the taste-breaking artifacts vanish in the continuum limit. Therefore, they cannot explain the deviation at the peak height, when only full (controlled continuum extrapolation) results are considered.

This points to an important advancement of the calculation described here over our previous results of~\cite{Borsanyi:2010cj}: we now include a large range of $N_t=12$ data points and one $N_t=16$ data point located at the peak position. Previously, we only had $N_t=12$ results at three characteristic temperature values available. Also, as mentioned above, the $T\rightarrow0$ limit is difficult due to taste-breaking effects, but is crucial since the renormalization is done at zero temperature, i.e. $p$($T$=0)=0. A mismatch at T=0 leads to a shift in the whole EoS. Previously, we calculated the difference in the pressure between the physical theory and its counterpart with 720~MeV heavy pions at a selected temperature ($100$~MeV), where the latter theory has practically zero pressure, and we, therefore, get $p(T=100~\textrm{MeV})$ in the physical theory with the desired normalization. The difference of this result and the prediction by the Hadron Resonance Gas model (HRG) was then included in the systematical error. With our increased range of temporal extents, we now can use five lattice spacings to fix the additive term in the pressure, arriving at a complete agreement with the hadron resonance gas model at low temperatures. We also improved the precision on our line of constant physics (LCP, see ref.~\cite{Borsanyi:2013bia}), and used two different methods to set the scale (based on the $w_0$ scale~\cite{Borsanyi:2012zs} or on $f_k$) in order to control the systematical error related to scale setting. 

These two different scale setting procedures entered into our 'histogram' method~\cite{Durr:2008zz} used to estimate systematical errors, along with a range of other fit methods, each of which is an in principle completely valid approach. We then calculated the goodness of fit Q and weights based on the Akaike information criterion AICc~\cite{AIC,AICc} and looked at the unweighted or weighted (based on Q or AICc) distribution of the results. The median is the central value, whereas the central region containing 68\% of all the possible methods gives an estimate on the systematic uncertainties. This procedure provides very conservative errors. Here, we had four basic types of continuum extrapolation methods (with or without tree level improvement for the pressure and with $a^2$ alone or $a^2$ and $a^4$ discretization effects) and two continuum extrapolation ranges (including or excluding the coarsest lattice $N_t$=6 in the analysis). We used seven ways to determine the subtraction term at T=0 (subtracting directly at the same gauge coupling $\beta$ or interpolating between the $\beta$ values with various orders of interpolation functions), and the aforementioned two scale procedures. Finally, we had eight options to determine the final trace anomaly by choosing among various spline functions, giving altogether 4$\cdot$2$\cdot$7$\cdot$2$\cdot$8=896 methods. Note that using either an AICc or Q based distribution changed the result only by a tiny fraction of the systematic uncertainty. Furthermore, the unweighted distribution always delivered consistent results within systematical errors.

The systematic error procedure clearly demonstrates the robustness of our final result. Even in the case of applying or not applying tree level improvement, where the data points at finite lattice spacing change considerably, the agreement between the continuum extrapolated results, and hence the contribution to the systematic error, is on the few percent level.

\section{Results}
We extended the ensembles available in refs.~\cite{Borsanyi:2010cj,Borsanyi:2013hza} by high precision (up to 67k trajectories) $T=0$ runs used for the LCP and subtraction and simulations on $32^3\times6$, $32^3\times8$ lattices, with $\sim$ 13k to 50k trajectories, and, to reduce the potential finite-volume effects, we also added six ensembles of $48^3\times 12$ lattices in the range $T=220\ldots335$~MeV with 30k trajectories, and $32^3\times 6$, $48^3\times 8$, $64^3\times 10$, and $64^3\times 12$, lattices with 5k, 40k, 10k, and 12k trajectories, respectively. 

To circumvent the algorithmic slowing down of the Hybrid Monte Carlo (HMC) algorithm at small lattice spacings, we chose to renormalize the higher temperatures ($T>355$~MeV) for the larger $N_t$ ensembles using the half-temperature subtraction described in refs.~\cite{Borsanyi:2012ve, Endrodi:2007tq}, which uses finite temperature ensembles in the deconfined phase (where the HMC still works) for the problematic lattice spacings. Here, we firstly subtract the value of the trace anomaly at the same coupling but doubled time extent (and thus a temperature of $T/2$ instead of $T$=0), i.e. $\left.(\varepsilon-3p)\right|_{T} -\left.(\varepsilon-3p)\right|_{T/2}$. Adding to this result the value of the trace anomaly at $T/2$ and the same $N_t$, we get the total trace anomaly. For the half-temperature subtractions we generated ensembles on $48^3\times 16$, $64^3\times 20$ and $64^3\times 24$ lattices with matching parameters and statistics to their finite temperature counterparts.

\begin{figure}
\begin{center}
\includegraphics*[width=7cm]{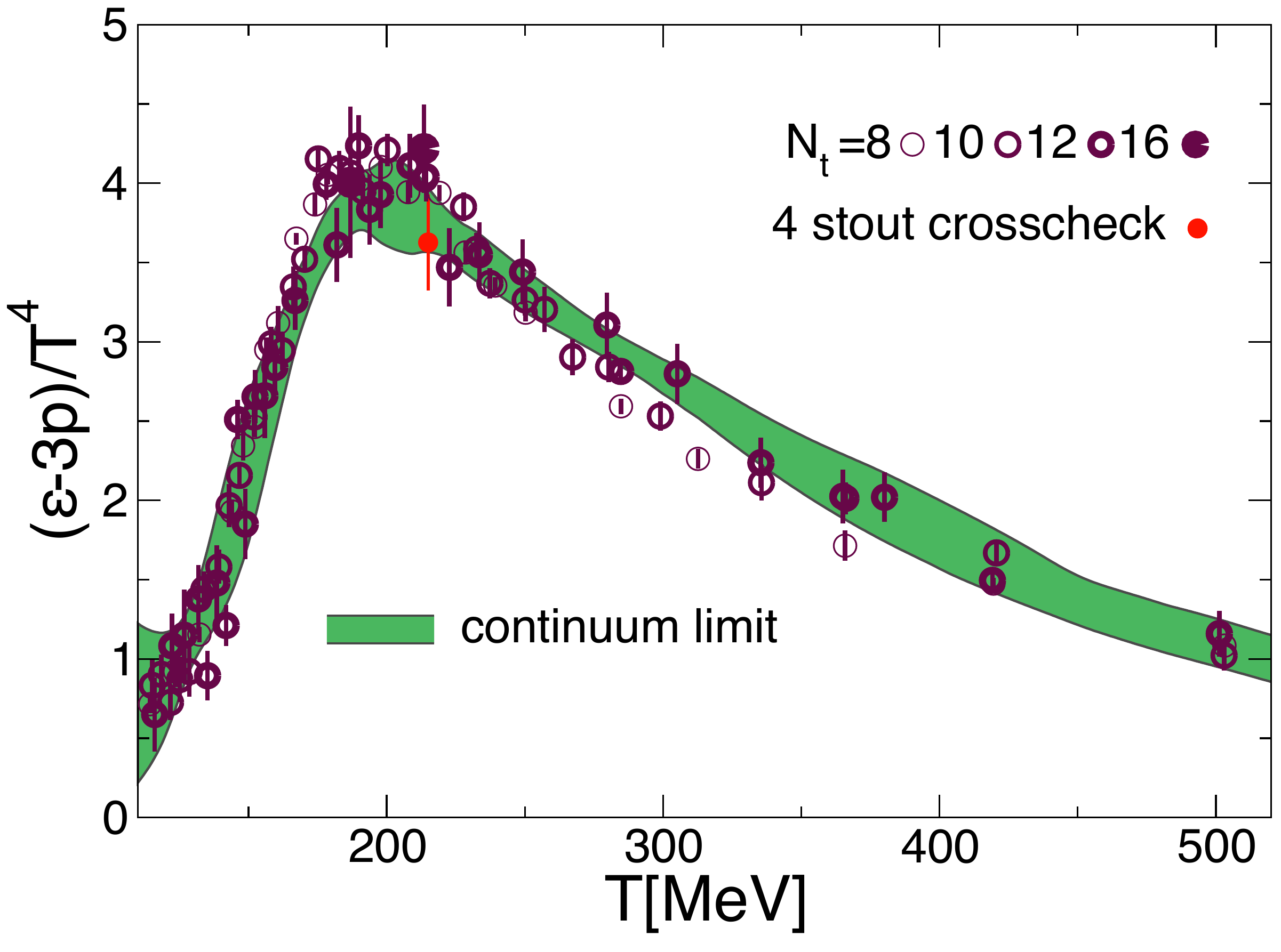}
\hspace{0.3cm}
\includegraphics*[width=7cm]{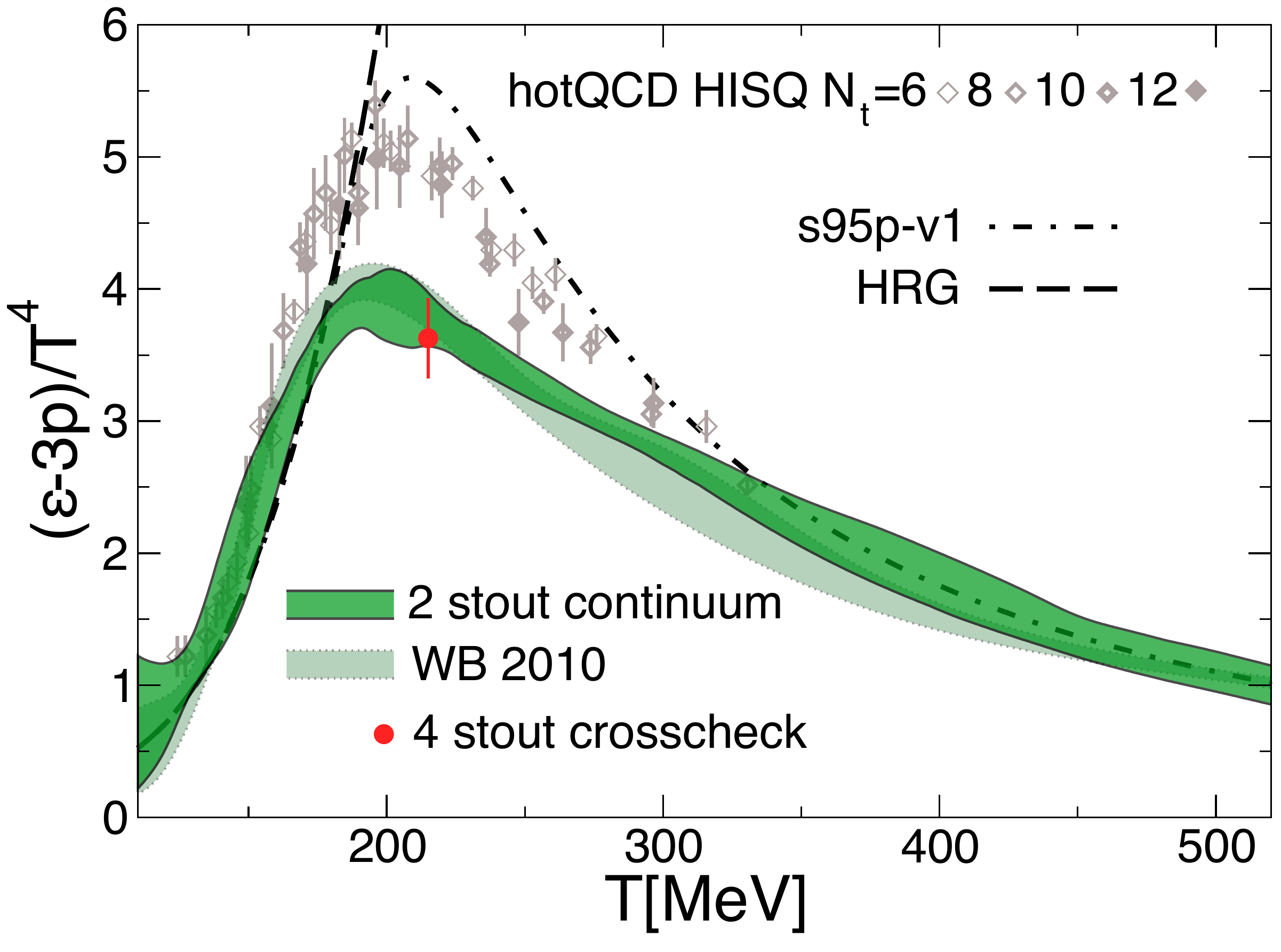}
\end{center}
\caption{\label{tracea_final}\textit{Left:}
the trace anomaly as a function of the temperature. The continuum extrapolated result with total errors is given by the shaded band. Also shown is a cross-check point computed in the continuum limit with a different action at $T=214$~MeV, indicated by a smaller filled red point, which serves as a crosscheck on the peak's hight (also on r.h.s.). 
\textit{Right:}
comparison of the result with HISQ results by the hotQCD collaboration (Lattice 2012~\cite{Petreczky:2012gi}, with $f_K$ scale setting) and the related parametrization 's95p-v1' of~\cite{Huovinen:2009yb}. A comparison to the Hadron Resonance Gas model's prediction and our result~\cite{Borsanyi:2010cj} from 2010 (``WB 2010'') is also shown.}
\end{figure}

The continuum extrapolated trace anomaly is shown in Figure~\ref{tracea_final} (left).
The results of the parallel investigation by the hotQCD group appear to be inconsistent with ours  (as of the lattice conference 2012~\cite{Petreczky:2012gi}). The situation might improve, when the HISQ analysis becomes complete with physical quark masses, a continuum extrapolation and a systematic error estimate.

The discrepancy visible in Figure~\ref{tracea_final} is most pronounced in the peak region, where we have (at $T\approx 214$~MeV) an $N_t=16$ data point in our continuum extrapolation. Using ensembles from our ongoing effort to compute the EoS of QCD with $N_f=2+1+1$ flavors, i.e. with dynamical charm quark, we added an additional continuum extrapolated cross-check point (see Figure~\ref{tracea_final}) at this same temperature (where the effect of the dynamical charm is not expected to be significant~\cite{Borsanyi:2012vn}). The action used in these calculations uses more smearing steps at a smaller smearing parameter than the one used in the $N_f=2+1$ calculations described here so far. The LCP was tuned completely independently by bracketing the physical point to $\pm 2\%$ in the quark masses, in boxes with $Lm_{\pi}>4$. The scale was set using the pion decay constant $f_\pi=130.41$~MeV (for further details see ref.~\cite{Borsanyi:2013bia}).

\begin{figure}
\begin{center}
\includegraphics*[width=7cm]{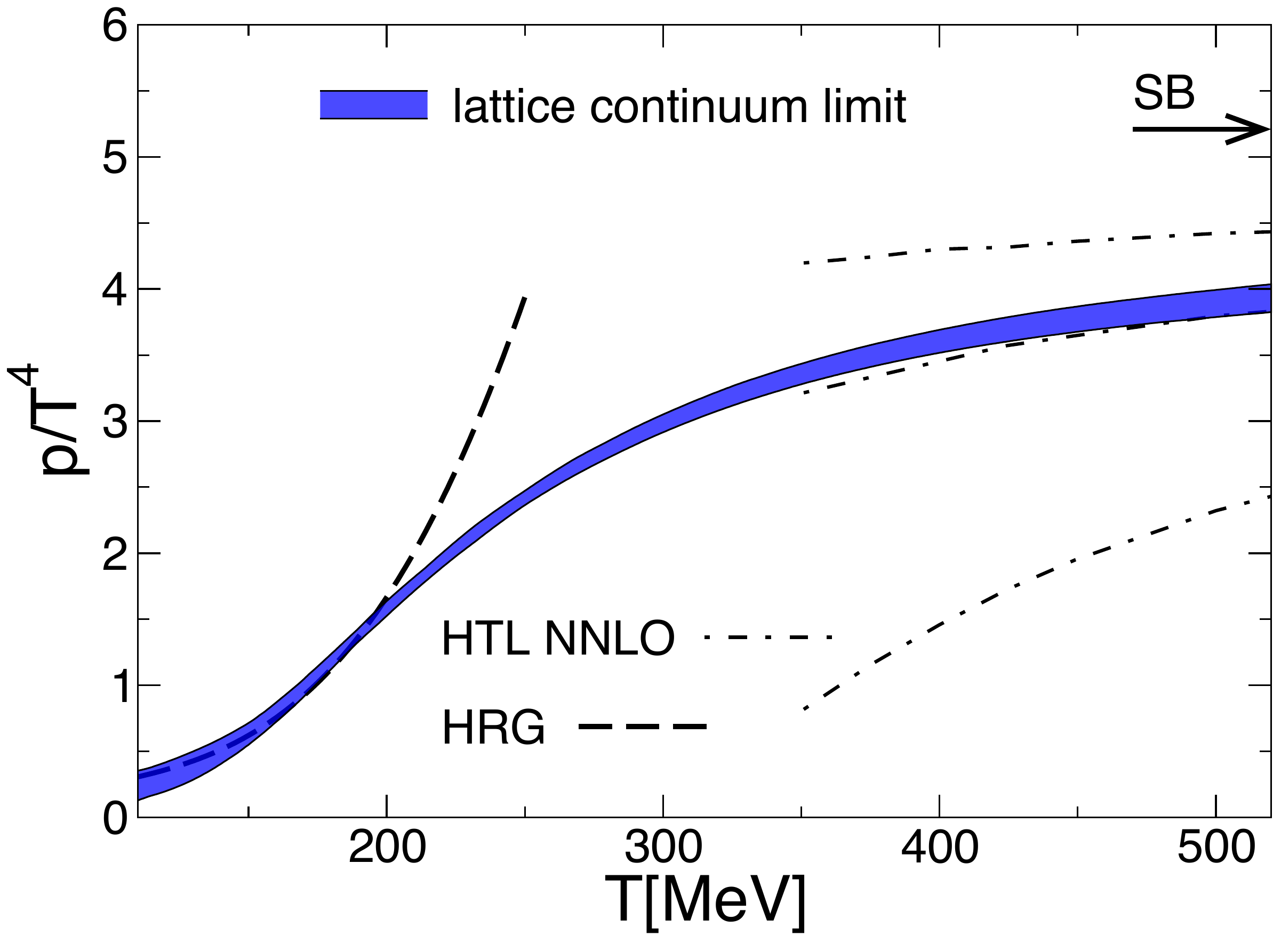}
\hspace{0.3cm}
\includegraphics*[width=7cm]{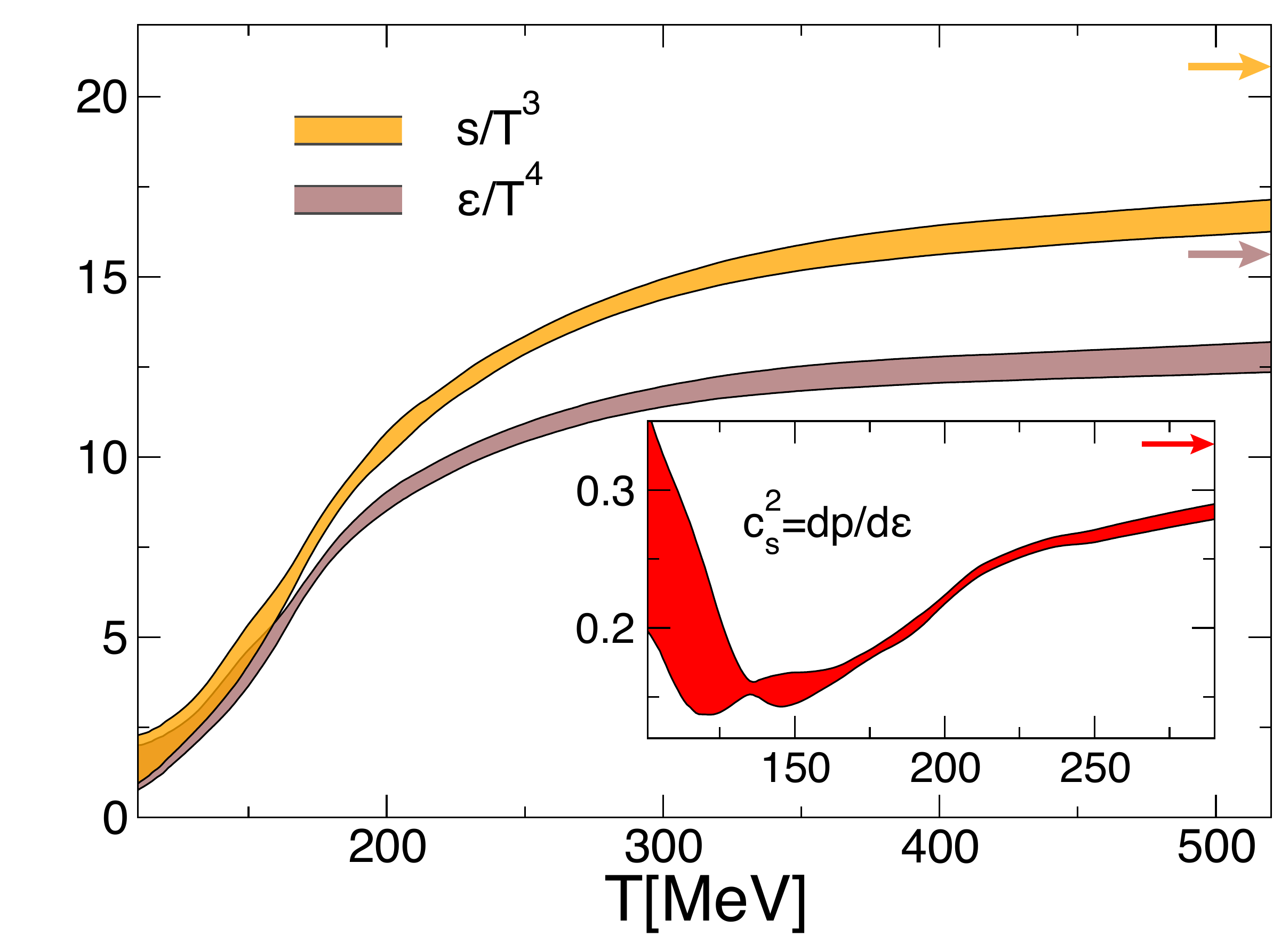}
\end{center}
\caption{\label{pressure}
{\em Left:}
continuum extrapolated result for the pressure with $N_f=2+1$ flavors, including 
the HRG prediction and a comparison to the NNLO Hard Thermal Loop
result of ref.~\cite{Andersen:2011sf} at high temperatures (renormalization
scales $\mu=$$\pi T$, $2\pi T$ or $4\pi T$).
{\em Right:} entropy and energy density. The insert shows the 
speed of sound.}
\end{figure}

The pressure is obtained via integration from the trace anomaly, see Figure~\ref{pressure} (left) together with the predictions of the hadron resonance gas (HRG) model at low temperatures. There
is a perfect agreement with HRG in the hadronic phase. The energy and entropy
densities as well as the speed of sound are shown in the right panel of Figure~\ref{pressure}.

The good agreement of these results with our 2010 ones~\cite{Borsanyi:2010cj}, with only a small variation in the high temperature region (T$>$350~MeV) where we now also use $N_t=10,12$ ensembles, prompts us to use the same functional form to parametrize our data:
\begin{equation}\label{fit_function} 
\frac{I(T)}{T^4} = \exp(-h_1/t-h_2/t^2)\cdot\left( h_0 + \frac{f_0 [ \cdot \tanh(f_1\cdot t+f_2)+1]}{1+g_1\cdot t + g_2 \cdot 
t^2} \right), 
\end{equation}
with slightly different fit parameters. Table~\ref{fit_parameters} contain the parametrization of ref.~\cite{Borsanyi:2010cj} and the parametrization of our present result. Note that though the two results differ only on the percent level, the parameters in the new parametrization changed more (these changes merely reflect some flat directions in the parameter space).

\begin{table}
\begin{center}
\begin{tabular}{|c||c|c|c|c|c|c|c|c|}
\hline
&$h_0$&$h_1$&$h_2$&$f_0$&$f_1$&$f_2$&$g_1$&$g_2$\\
\hline
\textbf{this work} & 0.1396 & -0.1800 & 0.0350 & 1.05 & 6.39 & -4.72 & -0.92 &0.57 \\
\textbf{2010~\cite{Borsanyi:2010cj}}& 0.1396 & -0.1800 & 0.0350 & 2.76 & 6.79 & -5.29 & -0.47 &1.04 \\
\hline
\end{tabular}
\end{center}
\caption{\label{fit_parameters}}
Constants for our parametrization of the trace anomaly in Eq.~(\ref{fit_function}).
\end{table}

\section{Conclusions}
We have presented a full result for the $N_f=2+1$ QCD equation of state. Our contiuum extrapolated results are completely consistent with our previous continuum estimate based on coarser lattices. The main advancement of the present work is the complete control over all systematic uncertainties.  We presented a parametrization of our result which makes it easy to use in other calculations and provide our tabulated results for download (see~\cite{resultsonline}).

\section*{Acknowledgments} Computations were performed on the Blue Gene 
supercomputer at FZ J\"ulich and on the QPACE machine and on GPU clusters 
\cite{Egri:2006zm} at University of Wuppertal. 
We acknowledge PRACE for awarding us resources on JUQUEEN at FZ J\"ulich.
CH wants to thank Utku~Can for interesting discussions.
This work was partially supported by the DFG Grant SFB/TRR 55. 
S. D. Katz is funded by the ERC grant ((FP7/2007-2013)/ERC No 208740) and the
Lend\"ulet program of the Hungarian Academy of Sciences ((LP2012-44/2012).


\begin{thebibliography}{99}

%\cite{Borsanyi:2013bia}
\bibitem{Borsanyi:2013bia} 
  S.~Borsanyi, Z.~Fodor, C.~Hoelbling, S.~D.~Katz, S.~Krieg and K.~K.~Szabo,
  %``Full result for the QCD equation of state with 2+1 flavors,''
  arXiv:1309.5258 [hep-lat].
  %%CITATION = ARXIV:1309.5258;%%

\bibitem{resultsonline}
The results are available as an ancillary file to \cite{Borsanyi:2013bia}, where they are freely downloadable. The file contains tabulated data for the trace anomaly, pressure, entropy, energy density, and speed of sound for our temperature range.

%\cite{Fodor:2009ax}
\bibitem{Fodor:2009ax}
  Z.~Fodor and S.~D.~Katz,
  %``The Phase diagram of quantum chromodynamics,''
  arXiv:0908.3341 [hep-ph].
  %%CITATION = ARXIV:0908.3341;%%
  %59 citations counted in INSPIRE as of 25 Jul 2013

%\cite{Philipsen:2012nu}
\bibitem{Philipsen:2012nu}
  O.~Philipsen,
  %``The QCD equation of state from the lattice,''
  Prog.\ Part.\ Nucl.\ Phys.\  {\bf 70} (2013) 55
  [arXiv:1207.5999 [hep-lat]].
  %%CITATION = ARXIV:1207.5999;%%
  %22 citations counted in INSPIRE as of 25 Jul 2013

%\cite{Petreczky:2013qj}
\bibitem{Petreczky:2013qj}
  P.~Petreczky,
  %``Review of recent highlights in lattice calculations at finite temperature and finite density,''
  PoS ConfinementX {\bf } (2012) 028
  [arXiv:1301.6188 [hep-lat]].
  %%CITATION = ARXIV:1301.6188;%%
  %1 citations counted in INSPIRE as of 25 Jul 2013

%\cite{Aoki:2006we}
\bibitem{Aoki:2006we}
  Y.~Aoki, G.~Endrodi, Z.~Fodor, S.~D.~Katz and K.~K.~Szabo,
  %``The Order of the quantum chromodynamics transition predicted by the standard model of particle physics,''
  Nature {\bf 443} (2006) 675
  [hep-lat/0611014].
  %%CITATION = HEP-LAT/0611014;%%
  %438 citations counted in INSPIRE as of 25 Jul 2013

%\cite{Aoki:2006br}
\bibitem{Aoki:2006br}
  Y.~Aoki, Z.~Fodor, S.~D.~Katz and K.~K.~Szabo,
  %``The QCD transition temperature: Results with physical masses in the continuum limit,''
  Phys.\ Lett.\ B {\bf 643} (2006) 46
  [hep-lat/0609068].
  %%CITATION = HEP-LAT/0609068;%%
  %424 citations counted in INSPIRE as of 25 Jul 2013

%\cite{Aoki:2009sc}
\bibitem{Aoki:2009sc}
  Y.~Aoki, S.~Borsanyi, S.~Durr, Z.~Fodor, S.~D.~Katz, S.~Krieg and K.~K.~Szabo,
  %``The QCD transition temperature: results with physical masses in the continuum limit II.,''
  JHEP {\bf 0906} (2009) 088
  [arXiv:0903.4155 [hep-lat]].
  %%CITATION = ARXIV:0903.4155;%%
  %265 citations counted in INSPIRE as of 25 Jul 2013

%\cite{Borsanyi:2010bp}
\bibitem{Borsanyi:2010bp}
  S.~Borsanyi {\it et al.}  [Wuppertal-Budapest Collaboration],
  %``Is there still any T_c mystery in lattice QCD? Results with physical masses in the continuum limit III,''
  JHEP {\bf 1009} (2010) 073
  [arXiv:1005.3508 [hep-lat]].
  %%CITATION = ARXIV:1005.3508;%%
  %246 citations counted in INSPIRE as of 25 Jul 2013

%\cite{Bazavov:2011nk}
\bibitem{Bazavov:2011nk}
  A.~Bazavov, T.~Bhattacharya, M.~Cheng, C.~DeTar, H.~T.~Ding, S.~Gottlieb, R.~Gupta and P.~Hegde {\it et al.},
  %``The chiral and deconfinement aspects of the QCD transition,''
  Phys.\ Rev.\ D {\bf 85} (2012) 054503
  [arXiv:1111.1710 [hep-lat]].
  %%CITATION = ARXIV:1111.1710;%%
  %150 citations counted in INSPIRE as of 25 Jul 2013

%\cite{Endrodi:2011gv}
\bibitem{Endrodi:2011gv}
  G.~Endrodi, Z.~Fodor, S.~D.~Katz and K.~K.~Szabo,
  %``The QCD phase diagram at nonzero quark density,''
  JHEP {\bf 1104} (2011) 001
  [arXiv:1102.1356 [hep-lat]].
  %%CITATION = ARXIV:1102.1356;%%
  %66 citations counted in INSPIRE as of 25 Jul 2013

%\cite{Borsanyi:2012cr}
\bibitem{Borsanyi:2012cr}
  S.~.Borsanyi, G.~Endrodi, Z.~Fodor, S.~D.~Katz, S.~Krieg, C.~Ratti and K.~K.~Szabo,
  %``QCD equation of state at nonzero chemical potential: continuum results with physical quark masses at order $mu^2$,''
  JHEP {\bf 1208} (2012) 053
  [arXiv:1204.6710 [hep-lat]].
  %%CITATION = ARXIV:1204.6710;%%
  %18 citations counted in INSPIRE as of 25 Jul 2013

%\cite{Fodor:2001au}
\bibitem{Fodor:2001au}
  Z.~Fodor and S.~D.~Katz,
  %``A New method to study lattice QCD at finite temperature and chemical potential,''
  Phys.\ Lett.\ B {\bf 534} (2002) 87
  [hep-lat/0104001].
  %%CITATION = HEP-LAT/0104001;%%
  %361 citations counted in INSPIRE as of 25 Jul 2013

%\cite{Petreczky:2012fsa}
\bibitem{Petreczky:2012fsa}
  P.~Petreczky,
  %``QCD at non-zero temperature: Status and prospects,''
  AIP Conf.\ Proc.\  {\bf 1520} (2013) 103.
  %%CITATION = APCPC,1520,103;%%

%\cite{Borsanyi:2010cj}
\bibitem{Borsanyi:2010cj}
  S.~Borsanyi, G.~Endrodi, Z.~Fodor, A.~Jakovac, S.~D.~Katz, S.~Krieg, C.~Ratti and K.~K.~Szabo,
  %``The QCD equation of state with dynamical quarks,''
  JHEP {\bf 1011} (2010) 077
  [arXiv:1007.2580 [hep-lat]].
  %%CITATION = ARXIV:1007.2580;%%
  %276 citations counted in INSPIRE as of 25 Jul 2013
  
%\cite{Aoki:2005vt}
\bibitem{Aoki:2005vt}
  Y.~Aoki, Z.~Fodor, S.~D.~Katz and K.~K.~Szabo,
  %``The Equation of state in lattice QCD: With physical quark masses towards the continuum limit,''
  JHEP {\bf 0601} (2006) 089
  [hep-lat/0510084].
  %%CITATION = HEP-LAT/0510084;%%
  %217 citations counted in INSPIRE as of 25 Jul 2013

%\cite{Peikert:1997vf}
\bibitem{Peikert:1997vf} 
  A.~Peikert, B.~Beinlich, A.~Bicker, F.~Karsch and E.~Laermann,
  %``Staggered fermion actions with improved rotational invariance,''
  Nucl.\ Phys.\ Proc.\ Suppl.\  {\bf 63}, 895 (1998)
  [hep-lat/9709157].
  %%CITATION = HEP-LAT/9709157;%%
  %13 citations counted in INSPIRE as of 16 Sep 2013  

%\cite{Borsanyi:2012zs}
\bibitem{Borsanyi:2012zs}
  S.~Borsanyi, S.~Durr, Z.~Fodor, C.~Hoelbling, S.~D.~Katz, S.~Krieg, T.~Kurth and L.~Lellouch {\it et al.},
  %``High-precision scale setting in lattice QCD,''
  JHEP {\bf 1209} (2012) 010
  [arXiv:1203.4469 [hep-lat]].
  %%CITATION = ARXIV:1203.4469;%%
  %12 citations counted in INSPIRE as of 25 Jul 2013

%\cite{Durr:2008zz}
\bibitem{Durr:2008zz}
  S.~Durr, Z.~Fodor, J.~Frison, C.~Hoelbling, R.~Hoffmann, S.~D.~Katz, S.~Krieg and T.~Kurth {\it et al.},
  %``Ab-Initio Determination of Light Hadron Masses,''
  Science {\bf 322} (2008) 1224
  [arXiv:0906.3599 [hep-lat]].
  %%CITATION = ARXIV:0906.3599;%%
  %245 citations counted in INSPIRE as of 25 Jul 2013
  
\bibitem{AIC}
 Hirotugu Akaike,
 IEEE Transactions on Automatic Control 19 (1974) 716-723
  
\bibitem{AICc}
 C.~M.~Hurvich, C.-L.~Tsai,
 Biometrika 76 (1989) 297-307

%\cite{Borsanyi:2013hza}
\bibitem{Borsanyi:2013hza} 
  S.~Borsanyi, Z.~Fodor, S.~D.~Katz, S.~Krieg, C.~Ratti and K.~K.~Szabo,
  %``Freeze-out parameters: lattice meets experiment,''
  Phys.\ Rev.\ Lett.\  {\bf 111}, 062005 (2013)
  [arXiv:1305.5161 [hep-lat]].
  %%CITATION = ARXIV:1305.5161;%%
  %11 citations counted in INSPIRE as of 04 Dec 2013  
  
%\cite{Borsanyi:2012ve}
\bibitem{Borsanyi:2012ve}
  S.~.Borsanyi, G.~Endrodi, Z.~Fodor, S.~D.~Katz and K.~K.~Szabo,
  %``Precision SU(3) lattice thermodynamics for a large temperature range,''
  JHEP {\bf 1207} (2012) 056
  [arXiv:1204.6184 [hep-lat]].
  %%CITATION = ARXIV:1204.6184;%%
  %24 citations counted in INSPIRE as of 25 Jul 2013

%\cite{Endrodi:2007tq}
\bibitem{Endrodi:2007tq}
  G.~Endrodi, Z.~Fodor, S.~D.~Katz and K.~K.~Szabo,
  %``The Equation of state at high temperatures from lattice QCD,''
  PoS LAT {\bf 2007} (2007) 228
  [arXiv:0710.4197 [hep-lat]].
  %%CITATION = ARXIV:0710.4197;%%
  %45 citations counted in INSPIRE as of 25 Jul 2013

%\cite{Petreczky:2012gi}
\bibitem{Petreczky:2012gi}
  P.~Petreczky [for HotQCD Collaboration],
  %``On trace anomaly in 2+1 flavor QCD,''
  PoS LATTICE {\bf 2012} (2012) 069
  [arXiv:1211.1678 [hep-lat]].
  %%CITATION = ARXIV:1211.1678;%%
  %4 citations counted in INSPIRE as of 25 Jul 2013

%\cite{Huovinen:2009yb}
\bibitem{Huovinen:2009yb} 
  P.~Huovinen and P.~Petreczky,
  %``QCD Equation of State and Hadron Resonance Gas,''
  Nucl.\ Phys.\ A {\bf 837}, 26 (2010)
  [arXiv:0912.2541 [hep-ph]].
  %%CITATION = ARXIV:0912.2541;%%
  %148 citations counted in INSPIRE as of 12 Aug 2013

%\cite{Borsanyi:2012vn}
\bibitem{Borsanyi:2012vn}
  S.~Borsanyi, G.~Endrodi, Z.~Fodor, S.~D.~Katz, S.~Krieg, C.~Ratti, C.~Schroeder and K.~K.~Szabo,
  %``The QCD equation of state and the effects of the charm,''
  PoS LATTICE {\bf 2011} (2011) 201
  [arXiv:1204.0995 [hep-lat]].
  %%CITATION = ARXIV:1204.0995;%%
  %7 citations counted in INSPIRE as of 25 Jul 2013

%\cite{Andersen:2011sf}
\bibitem{Andersen:2011sf}
  J.~O.~Andersen, L.~E.~Leganger, M.~Strickland and N.~Su,
  %``Three-loop HTL QCD thermodynamics,''
  JHEP {\bf 1108} (2011) 053
  [arXiv:1103.2528 [hep-ph]].
  %%CITATION = ARXIV:1103.2528;%%
  %29 citations counted in INSPIRE as of 25 Jul 2013

%\cite{Egri:2006zm}
\bibitem{Egri:2006zm}
  G.~I.~Egri, Z.~Fodor, C.~Hoelbling, S.~D.~Katz, D.~Nogradi and K.~K.~Szabo,
  %``Lattice QCD as a video game,''
  Comput.\ Phys.\ Commun.\  {\bf 177} (2007) 631
  [hep-lat/0611022].
  %%CITATION = HEP-LAT/0611022;%%
  %60 citations counted in INSPIRE as of 25 Jul 2013

\end{thebibliography}
\end{document}